\documentclass[journal=jcplcd,manuscript=letter]{achemso}

\usepackage[english]{babel}
\usepackage{chemformula} % Formula subscripts using \ch{}
\usepackage[T1]{fontenc} % Use modern font encodings

\usepackage{amsmath}
\usepackage{graphicx}
\usepackage[colorlinks=true, allcolors=blue]{hyperref}
\usepackage{siunitx}

\author{Ana Cadena}
\email{anacristina.cadena@wigner.hu}
\affiliation[Wigner RCP]
{Wigner Research Centre for Physics, Institute for Solid State Physics and Optics, 1525 Budapest, Hungary}
\alsoaffiliation{Department of Chemical and Environmental Process Engineering, Faculty of Chemical Technology and Biotechnology, Budapest University of Technology and Economics, 1111 Budapest, Hungary}
\author{Bea Botka}
\email{botka.bea@wigner.hu}
\author{\'Aron Pekker}
\affiliation[Wigner RCP]
{Wigner Research Centre for Physics, Institute for Solid State Physics and Optics, 1525 Budapest, Hungary}
\author{Cla Duri Tschannen}
\author{Chiara Lombardo }
\author{Lukas Novotny}
\affiliation[ETH]{Photonics Laboratory, ETH Z\"urich,
8093 Z\"urich, Switzerland}
\author{Andrei N. Khlobystov}
\affiliation[Nottingham]
{Department of Chemistry, University of Nottingham, Nottingham, NG7 2RD, United Kingdom}
\author{Katalin Kamar\'as}
\affiliation[Wigner RCP]
{Wigner Research Centre for Physics, Institute for Solid State Physics and Optics, 1525 Budapest, Hungary}

\title[Encapsulation from liquid phase]
{Molecular Encapsulation from the Liquid Phase and Graphene Nanoribbon Growth in Carbon Nanotubes}

\begin{document}

\begin{tocentry}
\includegraphics{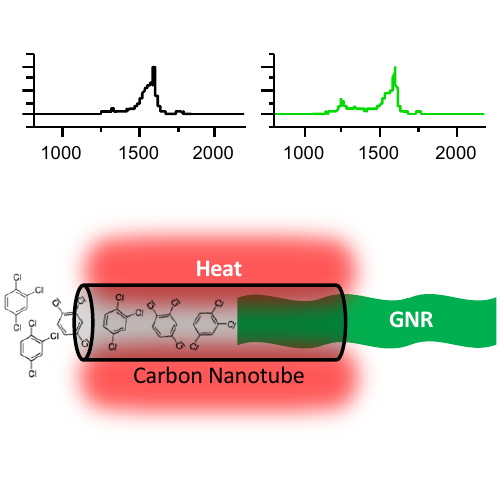}

\end{tocentry}

\makeatletter
\setlength\acs@tocentry@height{5.1cm}
\setlength\acs@tocentry@width{5.1cm}
\makeatother

\begin{abstract}

Growing graphene nanoribbons from small organic molecules encapsulated in carbon nanotubes can result in products with uniform width and chirality. We propose a method based on encapsulation of 1,2,4-trichlorobenzene from the liquid phase and subsequent annealing. This procedure results in graphene nanoribbons several tens of nanometers long. The presence of nanoribbons was proven by Raman spectra both on macroscopic samples and on the nanoscale by tip-enhanced Raman scattering and high-resolution transmission electron microscopic images.

\end{abstract}

Graphene nanoribbons (GNRs) are one-dimensional objects cut from graphene sheets with important applications in nanoelectronics. Their conformation can be zigzag (ZGNR), armchair (AGNR) or chiral. The ZGNRs are all predicted to be metallic, whereas AGNRs can be metallic or semiconducting, depending on their width. With this variety in size and conformation determining electronic properties, there arises a need for multiple fabrication methods, that are both selective and reproducible.

GNRs can be either produced from graphene by top-down methods or from small molecules by bottom-up procedures. \cite{Wang18} Top-down methods are various forms of lithography, or cutting by electron or ion beams or scanning probe tips.\cite{Tapaszto08}  Bottom-up procedures have been first performed in solution, for example branched polyphenylenes converted to highly conjugated graphite structures with high aspect ratio,\cite{Wu03} or 2D crystals  with long carbon chains stabilizing the edges.\cite{Yang08} Oriented chemical reactions on surfaces followed, starting from small planar molecules.\cite{Cai10,Houtsma21} Another possibility  is growing nanoribbons in a confined environment, where, besides the precursor, the container's internal diameter will determine the nanoribbon width. Carbon nanotubes (CNTs) are suitable templates for this purpose, because their diameter matches the size of the most popular nanoribbons 6-AGNR or 7-AGNR. Lim et al. \cite{Lim13} first suggested a two-step mechanism for the formation of inner nanotubes from encapsulated small molecules, where initial mild heat treatment produced GNRs that, on vigorous annealing, formed inner nanotubes through a twisted helical intermediate product. Besides leading to inner tubes of specific chirality, this study also suggested a method for producing encapsulated GNRs by stopping the process after the first step. Although intuition would dictate that fusion reactions of polyaromatic hydrocarbons would all follow this mechanism, this is not a general rule; only some specific reactions proceed through these intermediates, e.g. those starting from coronene \cite{Fujihara12,Botka14,Talyzin11} or perylene. Several other starting materials with different structure have been tried since, e.g. ferrocene,\cite{Kuzmany20,Zhang22} as well as precursors containing halogens \cite{Chamberlain17,Kinno19} and sulfur. \cite{Chuvilin11,Chamberlain12} These molecules show a surprising variety of structures, including non-six membered rings and functionalized fullerenes. The full understanding of the mechanism of this self-assembly into GNRs inside nanotubes is an open topic for experimental and theoretical efforts and underlines the need for further synthetic work to extend the selection of materials.

In this work we offer a new, facile method that can be applied to widen this database. We show that graphene nanoribbons with length several tens of nanometers can be grown from 1,2,4-trichlorobenzene (TCB) using a simple procedure. The encapsulation process results in a high filling ratio and does not require subsequent rinsing. We characterized the prepared GNRs, specifically the 6-AGNR, using Raman spectroscopy, tip-enhanced Raman scattering (TERS) and high-resolution transmission electron microscopy (TEM).

To obtain a nanoribbon with a controlled size and dimension, using CNTs as templates, it is very important to choose an appropriate encapsulation method. Such methods have been reviewed in our previous work.\cite{Cadena21} High temperature procedures can lead to the formation of undesired byproducts.\cite{Botka14} This can be particularly problematic in the case of GNR precursors, as GNR formation is generally initiated inside the nanotube by annealing. However, the molecules may start to react with each other already close to the sublimation temperature, leading to the outer walls of the nanotubes being covered with larger polycyclic aromatic hydrocarbons. Their removal can be problematic, as their solubility is rapidly decreasing with their size. Encapsulation from a solution of the precursor molecules\cite{Yudasaka03} overcomes this problem as it can be performed at low temperature and is suitable for a broad range of guest molecules. Its major disadvantage stems from the presence of the solvent that can also enter the nanotube channels. This not only leads to lower filling ratios compared to direct encapsulation methods, but can also alter the reaction pathway of the precursor inside the cavity.\cite{Miners16} Leaking of the guest molecules can also be problematic during rinsing steps, which are necessary to remove the non-encapsulated precursors. These steps may  decrease the filling ratio by diluting the solution in the cavity. An important development was direct immersion in neat organic liquids at room temperature, yielding complete filling of SWCNTs that was preserved in aqueous solutions.\cite{Campo16} A comprehensive study of encapsulated systems involving a wide range of nanotube diameters and guest molecules demonstrated the applicability of this method for optical property tuning.\cite{Campo20}

An ideal GNR precursor candidate would thus be a compound that is liquid at room temperature, therefore it can directly fill the immersed CNTs, and volatile enough, so the non-encapsulated molecules can be removed from the outer surface of the nanotubes by evaporation under mild enough conditions before forming ribbons inside. We found that TCB is suitable for this purpose.

Raman spectroscopy is the most widely employed method to detect subtle changes in nanotubes upon physical or chemical interaction with their environment.\cite{Dresselhaus10, Almadori14, Wenseleers07} Encapsulation of molecules can be easily detected by observing the shift of the radial breathing mode (RBM), as the molecules inside the nanotube cavity influence the radial expansion and contraction of the nanotube.\cite{Longhurst06} Data are available for a wide variety of encapsulated molecules.\cite{Cambre10,Campo20} In the case of individual nanotubes, a blueshift of the RBM occurs compared to the empty ones. A shift of the RBM can be also detected in bundles of tubes upon filling, though identifying the origin of the shift is more complicated in this case, as the interaction between the nanotubes also alters their RBM. In our work, after exposing the SWCNTs to TCB, a significant blueshift of the RBM mode (around 7 cm$^{-1}$) was detected with 532, 633 and 785 nm lasers (Figure S1, Supporting Information).

Raman spectra of GNRs can also be used for identification. Their most prominent feature is the radial breathing-like mode (RBLM),\cite{Zhou07} in the low-frequency part of the spectrum, corresponding to carbon atoms in the two halves of the ribbon moving in-phase in opposite directions. The frequency of this band depends on the ribbon width. As the edges of the ribbons are in most cases terminated by non-carbon atoms, typically hydrogen, C-H vibrations also appear, the most intense among them being the C-H in-plane bending (C-H ipb) mode. The Raman spectra of GNRs are resonant with the exciting laser, the excitation spectrum showing a maximum at their bandgap.

\begin{figure}[h!]
\centering
\includegraphics[width= 0.9 \linewidth]{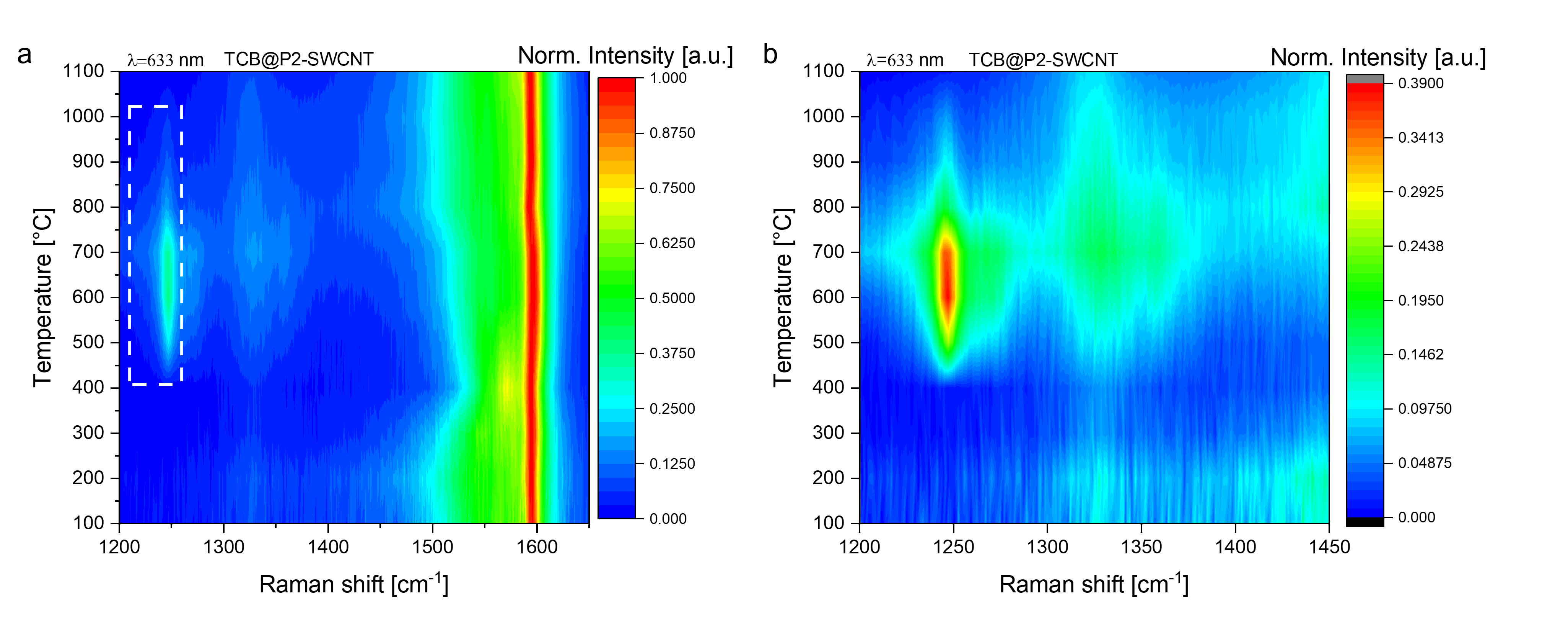}
\caption{Map representation of the Raman spectra taken during the annealing process. Each spectrum was collected at room temperature after consecutively annealing the sample at different temperatures (in 100 $^{\circ}$C steps) in a tube furnace for 12 hours. The sample was kept in vacuum in a closed quartz tube throughout the measurement sequence. The results were combined into a color coded map indicating GNR formation inside P2-SWCNT in a certain annealing temperature range. \textbf{a)} TCB@P2-SWCNT annealed from 100 to 1100 $^{\circ}$C in 100 $^{\circ}$C steps measured with a 633 nm laser through the quartz tube (the spectra were normalized to the SWCNT G mode). \textbf{b)} Close-up to the nanoribbon’s CH-ipb mode (marked by a rectangle) observed in map \textbf{a}.}
\label{Fig1}
\end{figure}

Encapsulation of TCB was performed by immersion of single-walled carbon nanotubes (P2, Carbon Solutions, Inc.\cite{carbsol}) in TCB at room temperature, resulting in the hybrid structure TCB@P2-SWCNT. To identify the ideal temperature at which GNRs form, we used a stepwise annealing protocol and followed the process by Raman spectroscopy. Starting from a cold furnace, the material was first annealed to 100 $^{\circ}$C for 12h and let cool to room temperature, then the Raman spectrum was recorded through the quartz tube; the procedure was repeated in 100 $^{\circ}$C steps up to 1100 $^{\circ}$C (Figure \ref{Fig1}). At this point the quartz tube was opened and the material measured under direct illumination, to identify double-walled carbon nanotubes whose RBM bands may be otherwise obscured by the quartz. Finally, the same sample was enclosed again in a quartz tube and annealed to 1200 $^{\circ}$C. The recorded Raman spectra are shown in Figure S2, Supporting Information. Based on the excitation profiles reported by Kuzmany et. al., \cite{Kuzmany20} a 633 nm laser can induce a resonant Raman process in 6-AGNRs. We used the CH in-plane bending (CH-ipb) mode to track the GNR formation during the annealing process (Figure \ref{Fig1}, Figure S2, Supporting Information). Formation of GNRs was first detected after annealing to 500 $^{\circ}$C (TCB@P2-SWCNT 500 $^{\circ}$C), with the most intense ribbon peaks appearing at 600 $^{\circ}$C formation temperature (TCB@P2-SWCNT 600 $^{\circ}$C). For formation temperatures above 700 $^{\circ}$C the ribbon modes start to diminish and they completely disappear at the 1100 $^{\circ}$C annealing step, while new peaks appear in the RBM region indicating the presence of inner nanotubes (Figure S2, Supporting Information). The conversion into inner tubes indirectly confirms  that the detected ribbons were formed in the nanotube interior.

It is useful to put our results in context with the literature. 6-AGNRs were observed to form from ferrocene at similar temperatures as from TCB. In the case of perylene, perylene-3,4,9,10-tetracarboxylic dianhydride and coronene, ribbon fragments started forming between 500 and 550 $^{\circ}$C, and GNRs appeared upon further annealing between 600 and 900  $^{\circ}$C. \cite{Lim13,Fujihara12, Botka14}  7-AGNRs from 10,10'-dibromo-9,9'-bianthryl were detected already after annealing at 300 $^{\circ}$C.\cite{Kinno19}

\begin{figure}[h!]
\centering
\includegraphics[width= 1 \linewidth]{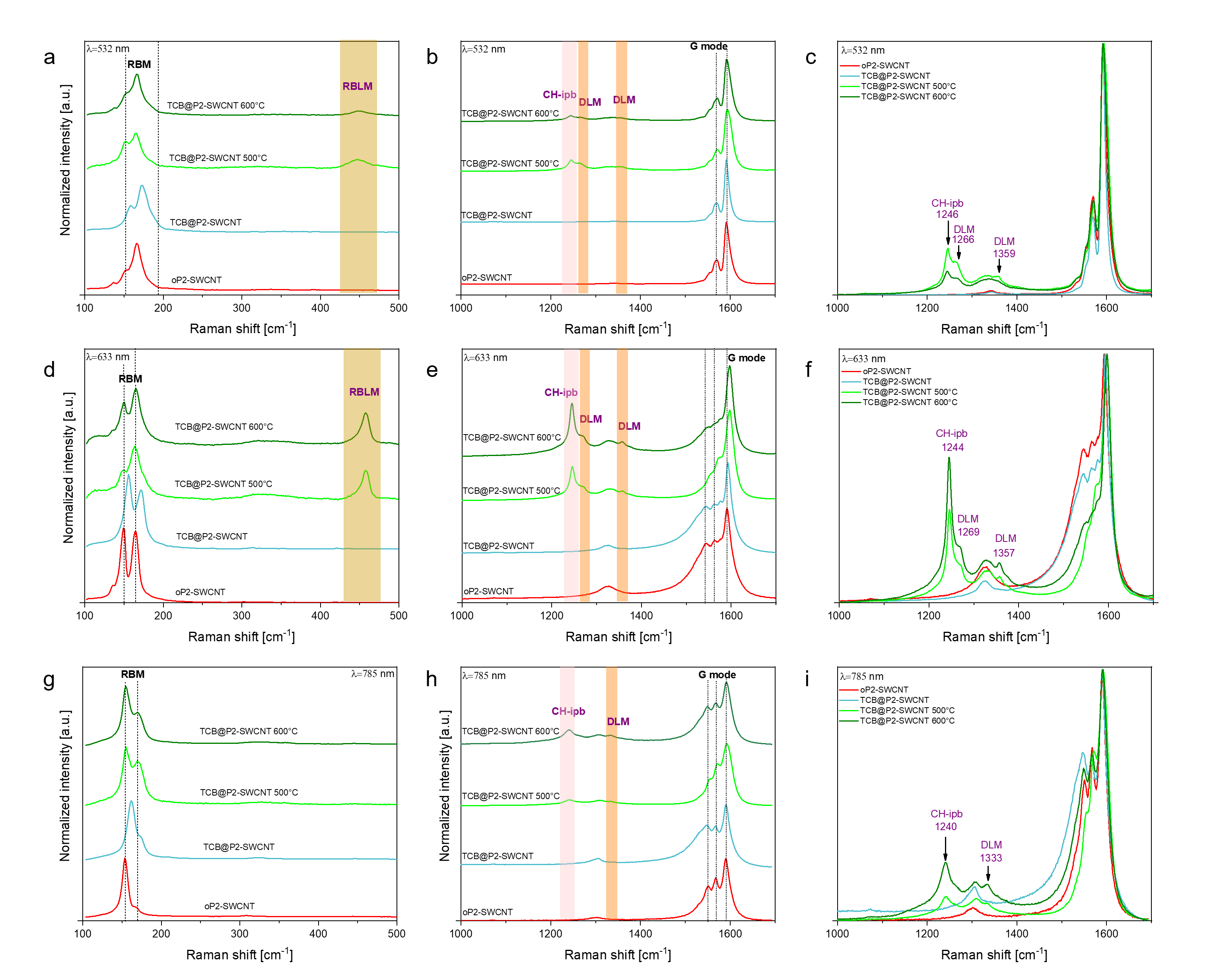}
\caption{Raman spectra of TCB@P2-SWCNT annealed to 500 and 600 $^{\circ}$C  compared with opened and filled P2-SWCNT as references. Spectra from the RBLM and CH-ipb-D regions: \textbf{a-c)} GNR@P2-SWCNT samples measured with 532 nm laser \textbf{d-f)} GNR@P2-SWCNT samples measured with 633 nm laser. \textbf{g-i)} GNR@P2-SWCNT samples measured with 785 nm laser. Low energy region spectra \textbf{a,d,g} are normalized to the RBM and \textbf{b,c,e,f,h,i} spectra to the G mode of the SWCNT.}
\label{Fig2}
\end{figure}

To perform a thorough characterization, the TCB@P2-SWCNT samples were directly annealed to the temperatures where the signal from the CH-ipb peak was the largest (500-600\ $^{\circ}$C). To prevent the encapsulated material from leaving the tubes during the low temperature annealing, the furnace was preheated to 500 $^{\circ}$C, then the sample was inserted and kept at this temperature for 12 hours. When the cooling step started, one end of the quartz tube was pulled out of the furnace to collect the molecules, if any, leaving the nanotube. The TCB@P2-SWCNT 600 $^{\circ}$C sample was prepared starting from the 500 $^{\circ}$C  sample following the same procedure. Inner tubes were formed by annealing the sample containing the ribbons (TCB@P2-SWCNT 600 $^{\circ}$C) to 1200 $^{\circ}$C (Figure S3, Supporting Information). Raman spectra were taken using 532, 633 and 785 nm excitation. Both the stepwise and the directly annealed samples were prepared from the same TCB@P2-SWCNT batch; judging the GNR yield by comparing the relative ratio of the ribbon peaks versus the CNT peaks, the directly annealed material shows higher conversion than the stepwise annealed one. The significant blueshift of the RBM observed on the TCB@P2-SWCNT is not present anymore after annealing (Figure \ref{Fig2}), indicating a relaxation of the tube walls from the strain exerted by the encapsulated TCB molecules. The latter may show a liquid-like association inside the tubes, similar to water filling,\cite{Cambre10} and convert to a less crowded occupancy of the interior by the ribbons.

It follows from the resonance character of the Raman spectra that excitation by different laser lines detects the presence of different types of both nanotubes and nanoribbons. First, we determine the possible combinations starting from the diameter distribution of the applied P2-SWCNTs:\cite{carbsol} 1.4 nm average, containing 1.2 -- 1.6 nm tubes,\cite{Pekker11} then we select the ribbons that fit into the given cavities. We follow the numbering scheme for nanoribbons presented by Fujita\cite{Fujita96} using the number of dimers (N) in the unit cell:  the atomic width indicates  the number of dimer lines in AGNRs and the number of zigzag lines in ZGNRs. Based on the observed RBM bands (Figure S1, Supporting Information), the three exciting lasers select the tubes in the 1.5 -- 1.6 nm diameter range.  We have calculated the width of the nanoribbons (solely the carbon network) by simply taking the distance between the outermost carbon atoms at the same position along the ribbon. The distances were calculated based on simple geometrical considerations giving negligible difference compared to the relaxed geometry published by Gillen et.al.\cite{Gillen09} (note the slightly different definition of width for ZGNR). The width was further increased by the projection of C-H bonds in the direction of the width. For the determination of the inner diameter of the smallest possible host carbon nanotubes, the carbon-hydrogen van der Waals distance was added to the width. These considerations lead to the possible nanoribbon types 4-7 AGNR and 3-4 ZGNR (see Figure S7 and Table S1, Supporting Information).

\begin{figure}[h!]
\centering
\includegraphics[width= 1 \linewidth]{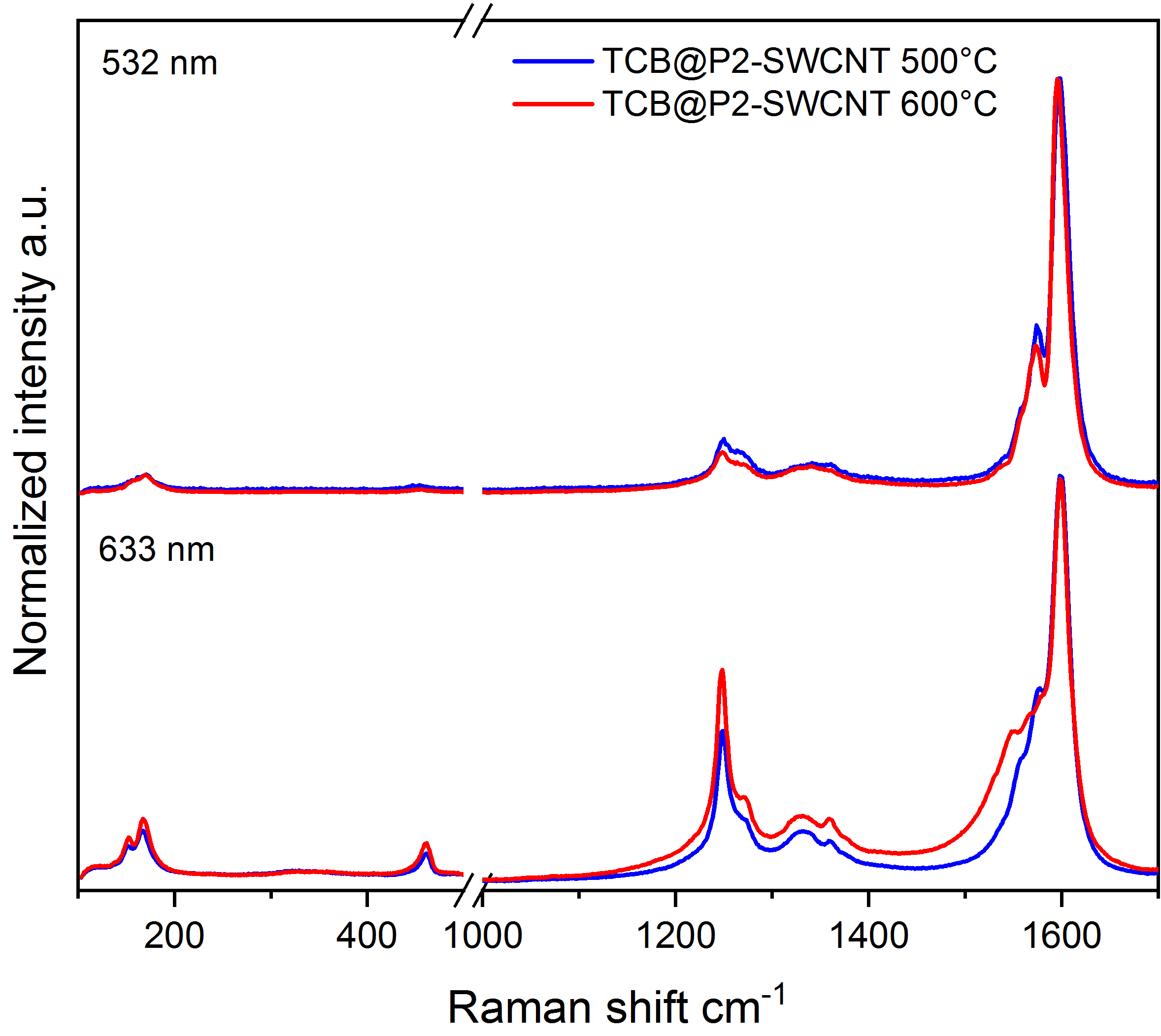}
\caption{GNR@P2-SWCNT formed at 500 $^{\circ}$C and 600 $^{\circ}$C measured with 532 and 633 nm laser excitation. GNR intensity ratio of the 500 and 600 $^{\circ}$C annealed sample changes with the excitation wavelength, indicating the presence of mostly shorter ribbons in the sample annealed at a lower temperature.}
\label{Fig3}
\end{figure}

 The RBLM of the 6-AGNR is clearly detected both with 633 and 532 nm excitation (Figure \ref{Fig2}a,d).  Its position is at 457 cm$^{-1}$ with 633 nm excitation, which is slightly redshifted compared to 6-AGNR@SWCNT prepared from the ferrocene precursor.\cite{Kuzmany20} Theoretical estimates for this mode are  457, \cite{Kuzmany20,Ma17}, 465.7 \cite{Zhou07} and $\sim$ 453 \cite{Liu20} cm$^{-1}$. Using 532 nm excitation, the RBLM is centered at 446 cm$^{-1}$. Since some nanotubes used in our experiment would also be of suitable size for the growth of either 5-AGNRs or 7-AGNRs, we searched for their signatures. We could neither observe the RBLM of the 7-AGNR at 397 cm$^{-1}$ with the  532 nm laser,\cite{Kinno19} nor that of the 5-AGNR at 534 cm$^{-1}$ with the 785 nm laser.\cite{Milotti22} Thus we can positively identify only the 6-AGNR, in accordance with TERS spectra (see below). As there are no experimental data for ZGNRs, we did not consider these conformations. As shown in Fig. \ref{Fig2}, the 6-AGNR modes are more intense in the 500 $^{\circ}$C  annealed sample using 532 nm excitation, while using 633 nm excitation the 600 $^{\circ}$C annealed sample has stronger GNR signals (Figure \ref{Fig3}). We presume this is caused by the presence of shorter ribbons at 500 $^{\circ}$C, which results in a blueshift of their bandgap. \cite{Talirz19}

Since we do not know the reaction mechanism of the ribbon formation, there is no \emph{a priori} knowledge of the range of the nanoribbons formed. Our case is more similar to the case of ferrocene, where the products could adapt to the diameter of the host tubes,\cite{Zhang22} than to that of larger polyaromatic molecules where the type of nanoribbon is pre-patterned by the precursors.\cite{Kinno19,Milotti22} Most probably, our samples show a heterogeneity in diameter, ribbon type, and edge termination. C-Cl bending vibrations, found around 200 cm$^{-1}$ in a TCB isomer,\cite{Saeki61} are not detected in the Raman spectra. Energy-dispersive X-ray (EDX) spectra (Fig. S5, Supporting Information), on the other hand, confirm the presence of Cl in the ribbons, indicating some chlorine termination, but quantitative information cannot be extracted. When estimating the possible fits of ribbon width and nanotube diameter, we have to allow for these ambiguities and also have to take into account that the hybrid structures are not exclusively the best tight fits.\cite{Kinno19} The nanotube RBM signals can therefore originate from a subset of filled nanotubes within a certain diameter distribution, as well as from empty nanotubes that fulfill the resonance condition. For this reason, we did not attempt to estimate a filling ratio or conversion yield from Raman spectra or compare it to other data in the literature.

\begin{figure}[h!]
\centering
\includegraphics[width= 1 \linewidth]{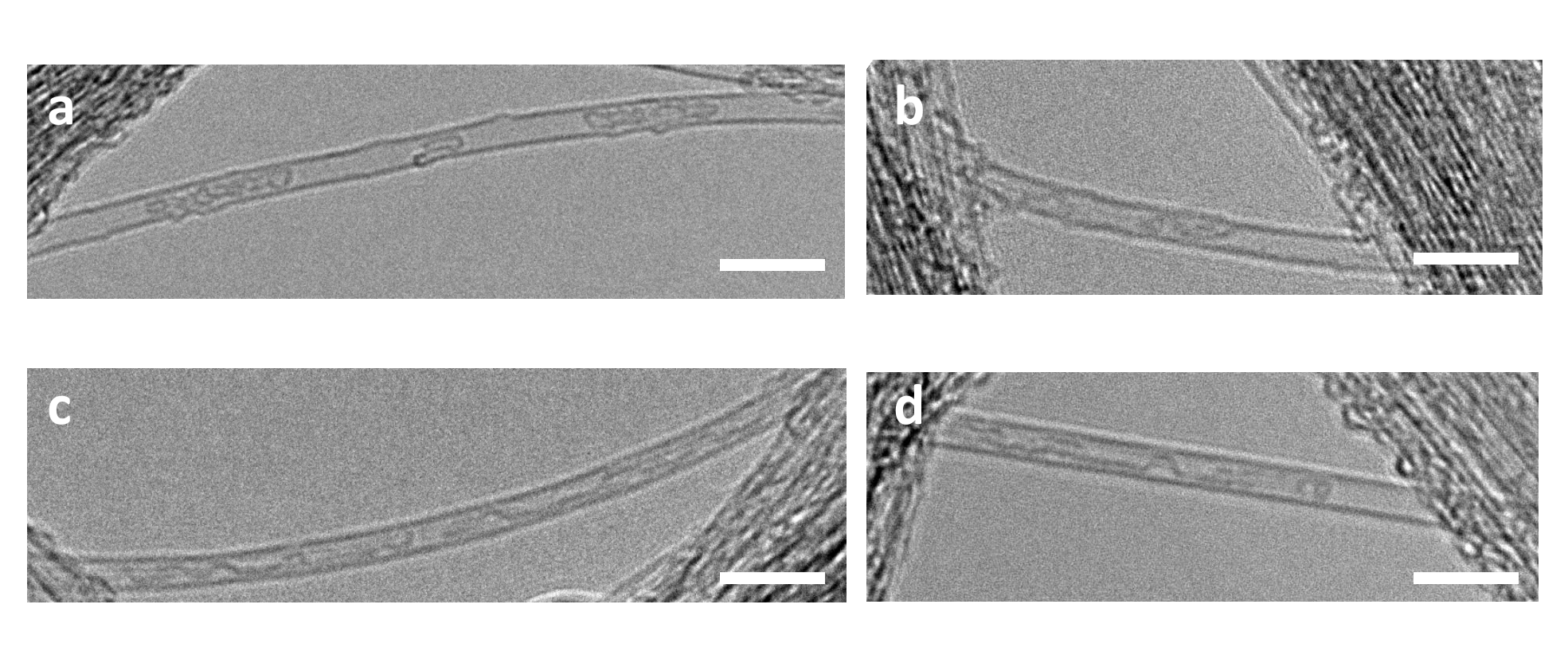}
\caption{Transmission electron microscopy images of the samples containing  graphene nanoribbons formed at \textbf{a-b)} 500 $^{\circ}$C, \textbf{c-d)} 600 $^{\circ}$C. The scalebar on the images is 10 nm.}
\label{Fig4}
\end{figure}

The products formed inside the carbon nanotubes can be visualized using transmission electron microscopy. The encapsulated nanoribbons are sensitive to the electron beam, they twist upon illumination, which can help identify these structures. \cite{Lim13, Chamberlain12} Such twisting ribbons can be observed both in the 500 $^{\circ}$C and 600 $^{\circ}$C annealed sample (Figure \ref{Fig4}). Confirming our spectroscopic observations, shorter ($\sim$ 20 nm) ribbons were observed in the sample annealed at 500 $^{\circ}$C than in the 600 $^{\circ}$C annealed one ($>$ 100 nm). TEM time sequences are shown in Figure S4, Supporting Information.

\begin{figure}[h!]
\centering
\includegraphics[width= 1 \linewidth]{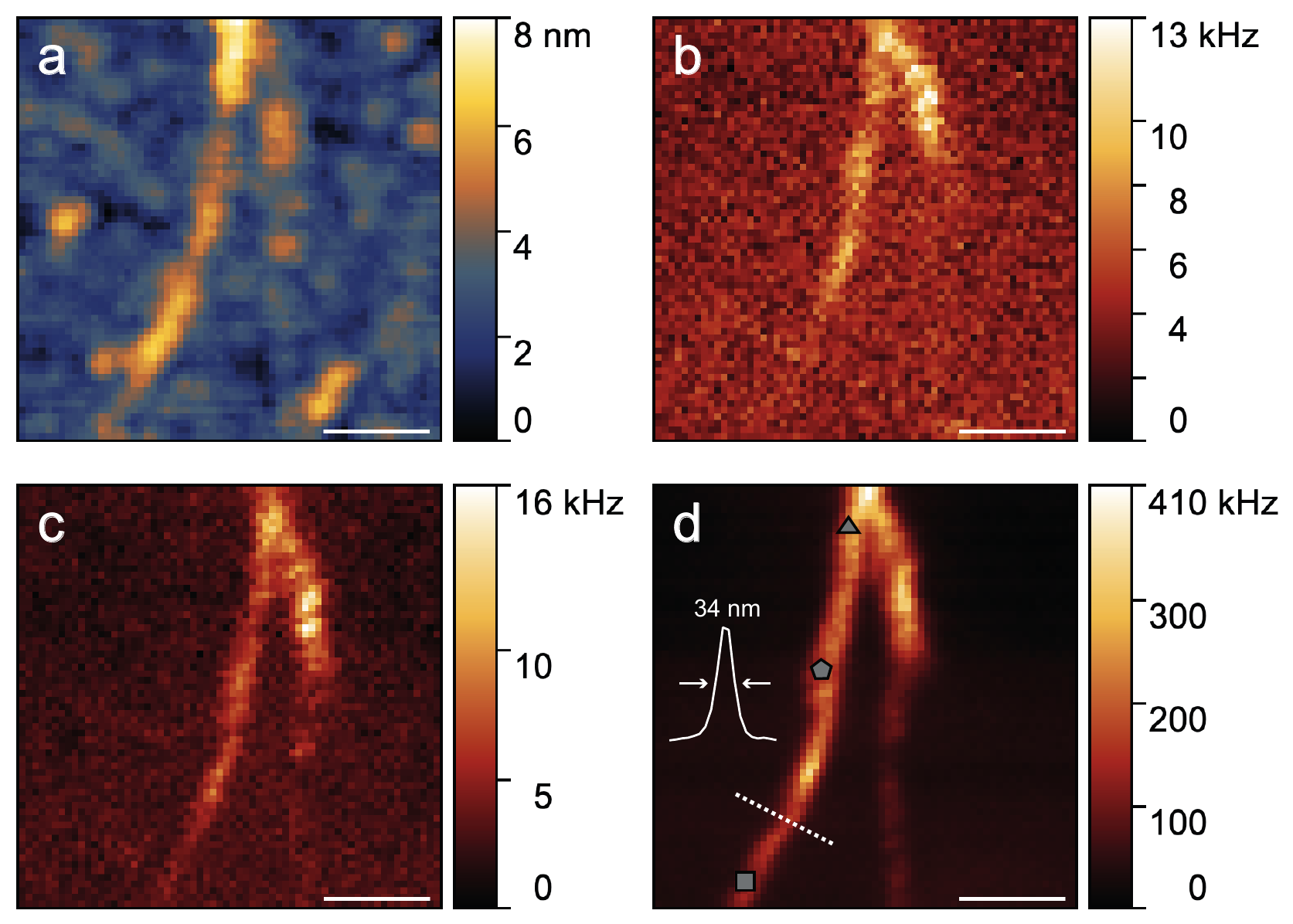}
\caption{Tip-enhanced Raman scattering of GNR@P2-SWCNT formed at \SI{600}{\celsius}. (a) Topographic image acquired simultaneously with the optical information. (b\textendash d) Optical image of the Stokes photons scattered by (b) the nanotube RBM, (c) the nanoribbon RBLM, and (d) the G mode of both nanotube and nanoribbon. The intensity profile in (d) extracted along the white dotted line indicates a spatial resolution of \SI{34}{\nano\metre}. TERS spectra recorded at the marked locations are shown in Fig.~S6, Supporting Information. The length of the scale bars corresponds to \SI{200}{\nano\metre}.}
\label{Fig5}
\end{figure}

To further verify that the spectral signatures observed in the far-field measurements presented in Figs.~\ref{Fig2} and \ref{Fig3} arise from encapsulated ribbons as identified by TEM in Fig.~\ref{Fig4}, we perform nanoscale imaging and spectroscopy using TERS. In Fig.~\ref{Fig5}, we plot several images of a small nanotube bundle depicting topographical [Fig.~\ref{Fig5}(a)] and optical [Fig.~\ref{Fig5}(b\textendash d)] information. The optical images in Fig.~\ref{Fig5}(b\textendash d) display the counts of Raman-scattered Stokes photons associated with the nanotube RBM, the ribbon RBLM, and the G mode from both nanotube and ribbon, respectively. Comparing these optical images clearly shows that the Raman signals arising from ribbon and nanotube are largely overlapping, indicating that ribbon formation occurs throughout the nanotubes and thus confirming the expected high filling ratio. TERS spectra recorded at the locations marked in Fig.~\ref{Fig5}(d) are shown in Fig.~S6, Supporting Information, along with the corresponding far-field measurements.

In conclusion, we report a method for growing graphene nanoribbons inside carbon nanotubes from 1,2,4-trichlorobenzene, a liquid at room temperature. 6-AGNRs were identified and characterized by Raman spectroscopy, TERS and TEM. The characteristic RBLM and CH-ipb modes of 6-AGNR were detected by far-field Raman spectroscopy and TERS. Following the Raman intensity of the CH-ipb peak it was possible to determine the ideal temperature for ribbon formation, and within that window, to follow the increase in length with increasing formation temperature. As the encapsulation step involves lower temperature than sublimation-based methods, and no additional solvent, it results in fewer potential byproducts and higher purity.

\section{Experimental methods}

The precleaned nanotubes were opened, degassed and immersed in
1,2,4-trichlorobenzene. The excess precursor was removed by filtration followed by evaporation of the solvent at room temperature.

\textbf {Encapsulating 1,2,4-trichlorobenzene into carbon nanotubes} P2-SWCNTs with 1.2-1.6 nm diameter range were purchased from Carbon Solutions, Inc.\cite{carbsol} TCB was purchased from Sigma Aldrich. Before filling, P2-SWCNTs were placed in a quartz tube and heated at 420 $^{\circ}$C  for 20 min in air to open the ends of the tubes. After removing the nanotube caps, functional groups that may block the open tube ends were removed by vacuum annealing. The tubes were placed under dynamic vacuum and inserted into a preheated furnace at 800~$^{\circ}$C  for 1 hour, then cooled down to ambient temperate in steps of 5 $^{\circ}$C/min. When the cleaned nanotubes reached room temperature, TCB was introduced into the quartz tube through a valve, without the nanotubes being exposed to air.  The mixture (TCB@P2-SWCNT) was sonicated for 10 min and filtered after 24 hours. To remove the molecules adsorbed on the walls, the TCB@P2-SWCNT was left open to air for four days.

\textbf{Preparation of GNR@P2-SWCNT} To determine the ideal temperature for ribbon formation, TCB@P2-SWCNT samples were annealed from 100 $^{\circ}$C to 1100 $^{\circ}$C  at 100 $^{\circ}$C  increments in a closed quartz tube. The annealing time at each temperature was kept for 12 hours, the heating and cooling rate was 5 $^{\circ}$C/min. After each annealing step the sample was analyzed using Raman spectroscopy through the quartz tube wall.
Once the ideal ribbon formation temperature was determined (Fig. \ref{Fig1}), another portion of the filled sample was annealed directly to 500 $^{\circ}$C and from there to 600 $^{\circ}$C. During the direct annealing, when the cooling down started, one end of the quartz tube was kept at a lower temperature, to allow condensation of desorbed guest species and reaction products, if any.

\textbf{Raman characterization} was carried out with a Renishaw InVia micro Raman spectrometer equipped with 532, 633 and 785 nm lasers. The measurements shown in Fig.~\ref{Fig2} were done using a sealed quartz ampoule. The samples were kept in the ampoule throughout the annealing sequence to prevent contamination. The micro Raman specra were obtained through the quartz using 20x magnification objectives. Regular measurements were performed in air with a 100x magnification objective. Laser power was kept low to prevent heating-induced shift of the Raman peaks and sample damage.

\textbf {Tip-Enhanced Raman Scattering} Samples for TERS were prepared by bath sonication in toluene (1 hour at 45 kHz 100 W followed by 30 min at 45 kHz 30 W). While still sonicating, a drop of the solution was pipetted onto the surface of distilled water and transferred to a glass coverslip with gold markers for TERS measurements.

TERS measurements are performed using the setup described in Ref.~\citenum{Tschannen22}. It consists of a home-built scanning probe microscope designed for non-contact shear-force operation with quartz tuning forks, on top of an inverted confocal microscope equipped with an $x,y$-scan stage. A radially polarized HeNe laser beam ($\lambda=633$\,\si{\nano\metre}) is strongly focused by a high-numerical-aperture (NA 1.4) oil-immersion objective through the thin glass coverslip carrying the sample. The TERS probe, a gold micropyramid attached to the end of a quartz tuning fork, is positioned into the center of the laser focus, thus generating a nanoscale excitation source for Raman scattering. The distance between the sample surface and the TERS probe is controlled by means of a shear-force feedback system. The backscattered light is collected by the same objective used for excitation and filtered by a long-pass filter (LP633) to remove the Rayleigh component. For imaging, the sample is raster-scanned while the tip--sample distance is held constant and the Raman-scattered light is collected by an avalanche photodiode (APD). The spectral region of interest is selected by appropriate optical filters placed in front of the APD. For the RBM [Fig.~\ref{Fig5}(b)], a band-pass filter centered at \SI{632}{\nano\metre} and with a full width at half maximum of \SI{22}{\nano\metre} (BP632/22) is used; the RBLM [Fig.~\ref{Fig5}(c)] is selected by a combination of LP647 and BP645/30; and for imaging of the G mode [Fig.~\ref{Fig5}(d)], we employ a BP632/22 filter. Full spectra (see Fig.~S6) are recorded using a CCD-equipped spectrometer. More details about the experimental implementation and the principles of TERS can be found in Ref.~\citenum{Tschannen22}.

\textbf {High resolution transmission electron microscopy} High-resolution TEM imaging was performed using a JEOL 2100F TEM (field emission gun source, information limit <0.19 nm) at 100 kV at room temperature. EDX spectra were recorded using an Oxford Instruments 30 mm2 Si(Li) detector or an Oxford Instruments x-Max 80 SDD running on an INCA microanalysis system. Samples for TEM and EDX were prepared by casting several drops of a suspension of nanotubes onto copper-grid mounted “lacey” carbon films.

%%%%%%%%%%%%%%%%%%%%%%%%%%%%%%%%%%%%%%%%%%%%%%%%%%%%%%%%%%%%%%%%%%%%%
\begin{acknowledgement}
We thank S\'andor Pekker and D\'avid F\"oldes for enlightening discussions.
Funding for this research was provided by the Hungarian National Research, Development and Innovation Office under grant no. FK 125063, FK 138411 and the Swiss National Science Foundation  under Grant No. $200020$\_$192362/1$. Research infrastructure in Hungary was provided by the Hungarian Academy of Sciences.
The authors thank the Nanoscale \& Microscale Research Centre (nmRC), University of Nottingham, for enabling access to HRTEM instrumentation.

\end{acknowledgement}

\begin{suppinfo}

The following files are available free of charge.
\begin{itemize}
  \item S1: additional Raman spectra, TEM time sequences, TERS spectra, AGNR width estimations
\end{itemize}

\end{suppinfo}

%\bibliography{Cadena-GNR}

\begin{mcitethebibliography}{38}
\providecommand*\natexlab[1]{#1}
\providecommand*\mciteSetBstSublistMode[1]{}
\providecommand*\mciteSetBstMaxWidthForm[2]{}
\providecommand*\mciteBstWouldAddEndPuncttrue
  {\def\EndOfBibitem{\unskip.}}
\providecommand*\mciteBstWouldAddEndPunctfalse
  {\let\EndOfBibitem\relax}
\providecommand*\mciteSetBstMidEndSepPunct[3]{}
\providecommand*\mciteSetBstSublistLabelBeginEnd[3]{}
\providecommand*\EndOfBibitem{}
\mciteSetBstSublistMode{f}
\mciteSetBstMaxWidthForm{subitem}{(\alph{mcitesubitemcount})}
\mciteSetBstSublistLabelBeginEnd
  {\mcitemaxwidthsubitemform\space}
  {\relax}
  {\relax}

\bibitem[Wang \latin{et~al.}(2018)Wang, Narita, and {M\"ullen}]{Wang18}
Wang,~X.-Y.; Narita,~A.; {M\"ullen},~K. Precision synthesis versus bulk-scale
  fabrication of graphenes. \emph{Nat. Rev. Chem.} \textbf{2018}, \emph{2},
  0100--1--10\relax
\mciteBstWouldAddEndPuncttrue
\mciteSetBstMidEndSepPunct{\mcitedefaultmidpunct}
{\mcitedefaultendpunct}{\mcitedefaultseppunct}\relax
\EndOfBibitem
\bibitem[Tapaszt\'o \latin{et~al.}(2008)Tapaszt\'o, Dobrik, Lambin, and
  Bir\'o]{Tapaszto08}
Tapaszt\'o,~L.; Dobrik,~G.; Lambin,~P.; Bir\'o,~L.~P. Tailoring the atomic
  structure of graphene nanoribbons by scanning tunneling microscope
  litography. \emph{Nat. Nanotechnol.} \textbf{2008}, \emph{3}, 397--401\relax
\mciteBstWouldAddEndPuncttrue
\mciteSetBstMidEndSepPunct{\mcitedefaultmidpunct}
{\mcitedefaultendpunct}{\mcitedefaultseppunct}\relax
\EndOfBibitem
\bibitem[Wu \latin{et~al.}(2003)Wu, Gherghel, Watson, Li, Wang, Simpson, Kolb,
  and {M\"ullen}]{Wu03}
Wu,~J.; Gherghel,~L.; Watson,~M.~D.; Li,~J.; Wang,~Z.; Simpson,~C.~D.;
  Kolb,~U.; {M\"ullen},~K. From branched polyphenylenes to graphite ribbons.
  \emph{Macromolecules} \textbf{2003}, \emph{36}, 7082--7089\relax
\mciteBstWouldAddEndPuncttrue
\mciteSetBstMidEndSepPunct{\mcitedefaultmidpunct}
{\mcitedefaultendpunct}{\mcitedefaultseppunct}\relax
\EndOfBibitem
\bibitem[Yang \latin{et~al.}(2008)Yang, Dou, Rouhanipour, Zhi, {R\"ader}, and
  {M\"ullen}]{Yang08}
Yang,~X.; Dou,~X.; Rouhanipour,~A.; Zhi,~L.; {R\"ader},~J.; {M\"ullen},~K.
  Two-dimensional graphene nanoribbons. \emph{J. Am. Chem. Soc.} \textbf{2008},
  \emph{130}, 4216--4217\relax
\mciteBstWouldAddEndPuncttrue
\mciteSetBstMidEndSepPunct{\mcitedefaultmidpunct}
{\mcitedefaultendpunct}{\mcitedefaultseppunct}\relax
\EndOfBibitem
\bibitem[Cai \latin{et~al.}(2010)Cai, Ruffieux, Jaafar, Bieri, Braun,
  Blankenburg, Muoth, Seitsonen, Saleh, Feng, Müllen, and Fasel]{Cai10}
Cai,~J.; Ruffieux,~P.; Jaafar,~R.; Bieri,~M.; Braun,~T.; Blankenburg,~S.;
  Muoth,~M.; Seitsonen,~A.; Saleh,~M.; Feng,~X.; Müllen,~K.; Fasel,~R.
  Atomically precise bottom-up fabrication of graphene nanoribbons.
  \emph{Nature} \textbf{2010}, \emph{466}, 470--473\relax
\mciteBstWouldAddEndPuncttrue
\mciteSetBstMidEndSepPunct{\mcitedefaultmidpunct}
{\mcitedefaultendpunct}{\mcitedefaultseppunct}\relax
\EndOfBibitem
\bibitem[Houtsma \latin{et~al.}(2021)Houtsma, {de la Rie}, and
  {St\"ohr}]{Houtsma21}
Houtsma,~R.; {de la Rie},~J.; {St\"ohr},~M. Atomically precise graphene
  nanoribbons: interplay of structural and electronic properties. \emph{Chem.
  Soc. Rev.} \textbf{2021}, \emph{50}, 6541--6568\relax
\mciteBstWouldAddEndPuncttrue
\mciteSetBstMidEndSepPunct{\mcitedefaultmidpunct}
{\mcitedefaultendpunct}{\mcitedefaultseppunct}\relax
\EndOfBibitem
\bibitem[Lim \latin{et~al.}(2013)Lim, Miyata, Kitaura, Nishimura, Nishimoto,
  Irle, Warner, Kataura, and Shinohara]{Lim13}
Lim,~H.~E.; Miyata,~Y.; Kitaura,~R.; Nishimura,~Y.; Nishimoto,~Y.; Irle,~S.;
  Warner,~J.~H.; Kataura,~H.; Shinohara,~H. Growth of carbon nanotubes via
  twisted graphene nanoribbons. \emph{Nat. Commun.} \textbf{2013}, \emph{4},
  2548--1--7\relax
\mciteBstWouldAddEndPuncttrue
\mciteSetBstMidEndSepPunct{\mcitedefaultmidpunct}
{\mcitedefaultendpunct}{\mcitedefaultseppunct}\relax
\EndOfBibitem
\bibitem[Fujihara \latin{et~al.}(2012)Fujihara, Miyata, Kitaura, Nishimura,
  Camacho, Irle, Iizumi, Okazaki, and Shinohara]{Fujihara12}
Fujihara,~M.; Miyata,~Y.; Kitaura,~R.; Nishimura,~O.; Camacho,~C.; Irle,~S.;
  Iizumi,~Y.; Okazaki,~T.; Shinohara,~H. Dimerization-initiated preferential
  formation of coronene-based graphene nanoribbons in carbon nanotubes.
  \emph{J. Phys. Chem. C} \textbf{2012}, \emph{116}, 15141--15145\relax
\mciteBstWouldAddEndPuncttrue
\mciteSetBstMidEndSepPunct{\mcitedefaultmidpunct}
{\mcitedefaultendpunct}{\mcitedefaultseppunct}\relax
\EndOfBibitem
\bibitem[Botka \latin{et~al.}(2014)Botka, {F{\"u}st{\"o}s}, T{\'o}h{\'a}ti,
  N{\'e}meth, Klupp, Szekr{\'e}nyes, Kocsis, Utcz{\'a}s, Sz{\'e}kely,
  V{\'a}czi, Tarczay, Hackl, Chamberlain, Khlobystov, and Kamar{\'a}s]{Botka14}
Botka,~B.; {F{\"u}st{\"o}s},~M.~E.; T{\'o}h{\'a}ti,~H.~M.; N{\'e}meth,~K.;
  Klupp,~G.; Szekr{\'e}nyes,~Z.; Kocsis,~D.; Utcz{\'a}s,~M.; Sz{\'e}kely,~E.;
  V{\'a}czi,~T.; Tarczay,~G.; Hackl,~R.; Chamberlain,~T.~W.; Khlobystov,~A.~N.;
  Kamar{\'a}s,~K. Interactions and chemical transformations of coronene inside
  and outside carbon nanotubes. \emph{Small} \textbf{2014}, \emph{10},
  1369--1378\relax
\mciteBstWouldAddEndPuncttrue
\mciteSetBstMidEndSepPunct{\mcitedefaultmidpunct}
{\mcitedefaultendpunct}{\mcitedefaultseppunct}\relax
\EndOfBibitem
\bibitem[Talyzin \latin{et~al.}(2011)Talyzin, Anoshkin, Krasheninnikov,
  Nieminen, Nasibulin, Jiang, and Kauppinen]{Talyzin11}
Talyzin,~A.~V.; Anoshkin,~I.~V.; Krasheninnikov,~A.~V.; Nieminen,~R.~M.;
  Nasibulin,~A.~G.; Jiang,~H.; Kauppinen,~E.~I. Synthesis of graphene
  nanoribbons encapsulated in single-walled carbon nanotubes. \emph{Nano Lett.}
  \textbf{2011}, \emph{11}, 4352--4356\relax
\mciteBstWouldAddEndPuncttrue
\mciteSetBstMidEndSepPunct{\mcitedefaultmidpunct}
{\mcitedefaultendpunct}{\mcitedefaultseppunct}\relax
\EndOfBibitem
\bibitem[Kuzmany \latin{et~al.}(2021)Kuzmany, Shi, Martinati, Cambr{\'e},
  Wenseleers, K{\"u}rti, Koltai, Kukucska, Cao, Kaiser, Saito, and
  Pichler]{Kuzmany20}
Kuzmany,~H.; Shi,~L.; Martinati,~M.; Cambr{\'e},~S.; Wenseleers,~W.;
  K{\"u}rti,~J.; Koltai,~J.; Kukucska,~G.; Cao,~K.; Kaiser,~U.; Saito,~T.;
  Pichler,~T. Well-defined sub-nanometer graphene ribbons synthesized inside
  carbon nanotubes. \emph{Carbon} \textbf{2021}, \emph{171}, 221--229\relax
\mciteBstWouldAddEndPuncttrue
\mciteSetBstMidEndSepPunct{\mcitedefaultmidpunct}
{\mcitedefaultendpunct}{\mcitedefaultseppunct}\relax
\EndOfBibitem
\bibitem[Zhang \latin{et~al.}(2022)Zhang, Chao, Saito, Kataura, Kuzmany,
  Pichler, Kaiser, Yang, and Shi]{Zhang22}
Zhang,~Y.; Chao,~K.; Saito,~T.; Kataura,~H.; Kuzmany,~H.; Pichler,~T.;
  Kaiser,~U.; Yang,~G.; Shi,~L. Carbon nanotube-dependent synthesis of armchair
  graphene nanoribbons. \emph{Nano Res.} \textbf{2022}, \emph{15},
  1709--1714\relax
\mciteBstWouldAddEndPuncttrue
\mciteSetBstMidEndSepPunct{\mcitedefaultmidpunct}
{\mcitedefaultendpunct}{\mcitedefaultseppunct}\relax
\EndOfBibitem
\bibitem[Chamberlain \latin{et~al.}(2017)Chamberlain, Biskupek, Skowron,
  Markevich, Kurasch, Reimer, Walker, Rance, Feng, {M\"ullen}, Turchanin,
  Lebedeva, Majouga, Nenajdenko, Kaiser, Besley, and Khlobystov]{Chamberlain17}
Chamberlain,~T.~W. \latin{et~al.}  Stop-frame filming and discovery of
  reactions at the single-molecule level by transmission electron microscopy.
  \emph{ACS Nano} \textbf{2017}, \emph{11}, 2509--2520\relax
\mciteBstWouldAddEndPuncttrue
\mciteSetBstMidEndSepPunct{\mcitedefaultmidpunct}
{\mcitedefaultendpunct}{\mcitedefaultseppunct}\relax
\EndOfBibitem
\bibitem[Kinno \latin{et~al.}(2019)Kinno, Omachi, and Shinohara]{Kinno19}
Kinno,~Y.; Omachi,~H.; Shinohara,~H. Template synthesis of armchair-edge
  graphene nanoribbons inside carbon nanotubes. \emph{Appl. Phys. Express}
  \textbf{2019}, \emph{13}, 015002--1--4\relax
\mciteBstWouldAddEndPuncttrue
\mciteSetBstMidEndSepPunct{\mcitedefaultmidpunct}
{\mcitedefaultendpunct}{\mcitedefaultseppunct}\relax
\EndOfBibitem
\bibitem[Chuvilin \latin{et~al.}(2011)Chuvilin, Bichoutskaia, Gimenez-Lopez,
  Chamberlain, Rance, Kuganathan, Biskupek, Kaiser, and Khlobystov]{Chuvilin11}
Chuvilin,~A.; Bichoutskaia,~E.; Gimenez-Lopez,~M.~C.; Chamberlain,~T.~W.;
  Rance,~G.~A.; Kuganathan,~N.; Biskupek,~J.; Kaiser,~U.; Khlobystov,~A.~N.
  Self-assembly of a sulphur-terminated graphene nanoribbon within a
  single-walled carbon nanotube. \emph{Nat. Mater.} \textbf{2011}, \emph{10},
  687--692\relax
\mciteBstWouldAddEndPuncttrue
\mciteSetBstMidEndSepPunct{\mcitedefaultmidpunct}
{\mcitedefaultendpunct}{\mcitedefaultseppunct}\relax
\EndOfBibitem
\bibitem[Chamberlain \latin{et~al.}(2012)Chamberlain, Biskupek, Rance,
  Chuvilin, Alexander, Bichoutskaia, Kaiser, and Khlobystov]{Chamberlain12}
Chamberlain,~T.~W.; Biskupek,~J.; Rance,~G.~A.; Chuvilin,~A.; Alexander,~T.~J.;
  Bichoutskaia,~E.; Kaiser,~U.; Khlobystov,~A.~N. Size, structure, and helical
  twist of graphene nanoribbons controlled by confinement in carbon nanotubes.
  \emph{ACS Nano} \textbf{2012}, \emph{6}, 3943--3953\relax
\mciteBstWouldAddEndPuncttrue
\mciteSetBstMidEndSepPunct{\mcitedefaultmidpunct}
{\mcitedefaultendpunct}{\mcitedefaultseppunct}\relax
\EndOfBibitem
\bibitem[Cadena \latin{et~al.}(2021)Cadena, Botka, and Kamar{\'a}s]{Cadena21}
Cadena,~A.; Botka,~B.; Kamar{\'a}s,~K. Organic molecules encapsulated in
  single-walled carbon nanotubes. \emph{Oxford Open Materials Science}
  \textbf{2021}, \emph{1}, itab009--1--16\relax
\mciteBstWouldAddEndPuncttrue
\mciteSetBstMidEndSepPunct{\mcitedefaultmidpunct}
{\mcitedefaultendpunct}{\mcitedefaultseppunct}\relax
\EndOfBibitem
\bibitem[Yudasaka \latin{et~al.}(2003)Yudasaka, Ajima, Suenaga, Ichihashi,
  Hashimoto, and Iijima]{Yudasaka03}
Yudasaka,~M.; Ajima,~K.; Suenaga,~K.; Ichihashi,~T.; Hashimoto,~A.; Iijima,~S.
  Nano-extraction and nano-condensation for {C$_{60}$} incorporation into
  single-wall carbon nanotubes in liquid phases. \emph{Chem. Phys. Lett.}
  \textbf{2003}, \emph{380}, 42--46\relax
\mciteBstWouldAddEndPuncttrue
\mciteSetBstMidEndSepPunct{\mcitedefaultmidpunct}
{\mcitedefaultendpunct}{\mcitedefaultseppunct}\relax
\EndOfBibitem
\bibitem[Miners \latin{et~al.}(2016)Miners, Rance, and Khlobystov]{Miners16}
Miners,~S.~A.; Rance,~G.~A.; Khlobystov,~A.~N. Chemical reactions confined
  within carbon nanotubes. \emph{Chem. Soc. Rev.} \textbf{2016}, \emph{45},
  4727--4746\relax
\mciteBstWouldAddEndPuncttrue
\mciteSetBstMidEndSepPunct{\mcitedefaultmidpunct}
{\mcitedefaultendpunct}{\mcitedefaultseppunct}\relax
\EndOfBibitem
\bibitem[Campo \latin{et~al.}(2016)Campo, Piao, Lam, Stafford, Streit, Simpson,
  {Hight Walker}, and Fagan]{Campo16}
Campo,~J.; Piao,~Y.; Lam,~S.; Stafford,~C.; Streit,~J.; Simpson,~J.; {Hight
  Walker},~A.; Fagan,~J. Enhancing single-Wall carbon nanotube properties
  through controlled endohedral filling. \emph{Nanoscale Horiz.} \textbf{2016},
  \emph{1}, 317--324\relax
\mciteBstWouldAddEndPuncttrue
\mciteSetBstMidEndSepPunct{\mcitedefaultmidpunct}
{\mcitedefaultendpunct}{\mcitedefaultseppunct}\relax
\EndOfBibitem
\bibitem[Campo \latin{et~al.}(2021)Campo, Cambr{\'e}, Botka, Obrzut,
  Wenseleers, and Fagan]{Campo20}
Campo,~J.; Cambr{\'e},~S.; Botka,~B.; Obrzut,~J.; Wenseleers,~W.; Fagan,~J.
  Optical property tuning of single-wall carbon nanotubes by endohedral
  encapsulation of a wide variety of dielectric molecules. \emph{ACS Nano}
  \textbf{2021}, \emph{15}, 2301--2317\relax
\mciteBstWouldAddEndPuncttrue
\mciteSetBstMidEndSepPunct{\mcitedefaultmidpunct}
{\mcitedefaultendpunct}{\mcitedefaultseppunct}\relax
\EndOfBibitem
\bibitem[Dresselhaus \latin{et~al.}(2010)Dresselhaus, Jorio, Hofmann,
  Dresselhaus, and Saito]{Dresselhaus10}
Dresselhaus,~M.; Jorio,~A.; Hofmann,~M.; Dresselhaus,~G.; Saito,~R.
  Perspectives on carbon nanotubes and graphene Raman spectroscopy. \emph{Nano
  Lett.} \textbf{2010}, \emph{3}, 751--758\relax
\mciteBstWouldAddEndPuncttrue
\mciteSetBstMidEndSepPunct{\mcitedefaultmidpunct}
{\mcitedefaultendpunct}{\mcitedefaultseppunct}\relax
\EndOfBibitem
\bibitem[Almadori \latin{et~al.}(2014)Almadori, Alvarez, {Le Parc}, Aznar,
  Fossard, Loiseau, Jousselme, Campidelli, Hermet, Belhboub, Rahmani, Saito,
  and Bantignies]{Almadori14}
Almadori,~Y.; Alvarez,~L.; {Le Parc},~R.; Aznar,~R.; Fossard,~F.; Loiseau,~A.;
  Jousselme,~B.; Campidelli,~S.; Hermet,~P.; Belhboub,~A.; Rahmani,~A.;
  Saito,~T.; Bantignies,~J.-L. Chromophore ordering by confinement into carbon
  nanotubes. \emph{J. Phys. Chem. C} \textbf{2014}, \emph{118},
  19462--19468\relax
\mciteBstWouldAddEndPuncttrue
\mciteSetBstMidEndSepPunct{\mcitedefaultmidpunct}
{\mcitedefaultendpunct}{\mcitedefaultseppunct}\relax
\EndOfBibitem
\bibitem[Wenseleers \latin{et~al.}(2007)Wenseleers, Cambr{\'e}, {\v{C}}ulin,
  Bouwen, and Goovaerts]{Wenseleers07}
Wenseleers,~W.; Cambr{\'e},~S.; {\v{C}}ulin,~J.; Bouwen,~A.; Goovaerts,~E.
  Effect of water filling on the electronic and vibrational resonances of
  carbon nanotubes: Characterizing tube opening by Raman spectroscopy.
  \emph{Adv. Mater.} \textbf{2007}, \emph{19}, 2274--2278\relax
\mciteBstWouldAddEndPuncttrue
\mciteSetBstMidEndSepPunct{\mcitedefaultmidpunct}
{\mcitedefaultendpunct}{\mcitedefaultseppunct}\relax
\EndOfBibitem
\bibitem[Longhurst and Quirke(2006)Longhurst, and Quirke]{Longhurst06}
Longhurst,~M.~J.; Quirke,~N. The environmental effect on the radial breathing
  mode of carbon nanotubes. II. Shell model approximation for internally and
  externally adsorbed fluids. \emph{J. Chem. Phys.} \textbf{2006}, \emph{125},
  184705--1--6\relax
\mciteBstWouldAddEndPuncttrue
\mciteSetBstMidEndSepPunct{\mcitedefaultmidpunct}
{\mcitedefaultendpunct}{\mcitedefaultseppunct}\relax
\EndOfBibitem
\bibitem[Cambr{\'e} \latin{et~al.}(2010)Cambr{\'e}, Schoeters, Luyckx,
  Goovaerts, and Wenseleers]{Cambre10}
Cambr{\'e},~S.; Schoeters,~B.; Luyckx,~S.; Goovaerts,~E.; Wenseleers,~W.
  Experimental observation of single-file water filling of thin single-wall
  carbon nanotubes down to chiral index (5,3). \emph{Phys. Rev. Lett}
  \textbf{2010}, \emph{104}, 207401--1--4\relax
\mciteBstWouldAddEndPuncttrue
\mciteSetBstMidEndSepPunct{\mcitedefaultmidpunct}
{\mcitedefaultendpunct}{\mcitedefaultseppunct}\relax
\EndOfBibitem
\bibitem[Zhou and Dong(2007)Zhou, and Dong]{Zhou07}
Zhou,~J.; Dong,~J. Vibrational property and Raman spectrum of carbon
  nanoribbon. \emph{Appl. Phys. Lett.} \textbf{2007}, \emph{91},
  173108--1--3\relax
\mciteBstWouldAddEndPuncttrue
\mciteSetBstMidEndSepPunct{\mcitedefaultmidpunct}
{\mcitedefaultendpunct}{\mcitedefaultseppunct}\relax
\EndOfBibitem
\bibitem[car()]{carbsol}
Carbon {Solutions}, {Inc.} \url{http://www.carbonsolution.com}, Accessed:
  2022-08-15\relax
\mciteBstWouldAddEndPuncttrue
\mciteSetBstMidEndSepPunct{\mcitedefaultmidpunct}
{\mcitedefaultendpunct}{\mcitedefaultseppunct}\relax
\EndOfBibitem
\bibitem[Pekker and Kamar\'as(2011)Pekker, and Kamar\'as]{Pekker11}
Pekker,~{\'A}.; Kamar\'as,~K. Wide-range optical studies on various
  single-walled carbon nanotubes: origin of the low-energy gap. \emph{Phys.
  Rev. B} \textbf{2011}, \emph{84}, 075475\relax
\mciteBstWouldAddEndPuncttrue
\mciteSetBstMidEndSepPunct{\mcitedefaultmidpunct}
{\mcitedefaultendpunct}{\mcitedefaultseppunct}\relax
\EndOfBibitem
\bibitem[Fujita \latin{et~al.}(1996)Fujita, Wakabayashi, Nakada, and
  Kusakabe]{Fujita96}
Fujita,~M.; Wakabayashi,~K.; Nakada,~K.; Kusakabe,~K. Peculiar localized state
  at zigzag graphite edge. \emph{J. Phys. Soc. Jpn.} \textbf{1996}, \emph{65},
  1920--1923\relax
\mciteBstWouldAddEndPuncttrue
\mciteSetBstMidEndSepPunct{\mcitedefaultmidpunct}
{\mcitedefaultendpunct}{\mcitedefaultseppunct}\relax
\EndOfBibitem
\bibitem[Gillen \latin{et~al.}(2009)Gillen, Mohr, Thomsen, and
  Maultzsch]{Gillen09}
Gillen,~R.; Mohr,~M.; Thomsen,~C.; Maultzsch,~J. Vibrational properties of
  graphene nanoribbons by first-principles calculations. \emph{Phys. Rev. B}
  \textbf{2009}, \emph{80}, 155418\relax
\mciteBstWouldAddEndPuncttrue
\mciteSetBstMidEndSepPunct{\mcitedefaultmidpunct}
{\mcitedefaultendpunct}{\mcitedefaultseppunct}\relax
\EndOfBibitem
\bibitem[Ma \latin{et~al.}(2017)Ma, Liang, Xiao, Puretzky, Hong, Lu, Meunier,
  Bernholc, and Li]{Ma17}
Ma,~C.; Liang,~L.; Xiao,~Z.; Puretzky,~A.~A.; Hong,~K.; Lu,~W.; Meunier,~V.;
  Bernholc,~J.; Li,~A.-P. Seamless staircase electrical contact to
  semiconducting graphene nanoribbons. \emph{Nano Lett.} \textbf{2017},
  \emph{17}, 6241--6247\relax
\mciteBstWouldAddEndPuncttrue
\mciteSetBstMidEndSepPunct{\mcitedefaultmidpunct}
{\mcitedefaultendpunct}{\mcitedefaultseppunct}\relax
\EndOfBibitem
\bibitem[Liu \latin{et~al.}(2020)Liu, Daniels, Meunier, Every, and
  Tom{\'a}nek]{Liu20}
Liu,~D.; Daniels,~C.; Meunier,~V.; Every,~A.~G.; Tom{\'a}nek,~D. In-plane
  breathing and shear modes in low-dimensional nanostructures. \emph{Carbon}
  \textbf{2020}, \emph{157}, 364--370\relax
\mciteBstWouldAddEndPuncttrue
\mciteSetBstMidEndSepPunct{\mcitedefaultmidpunct}
{\mcitedefaultendpunct}{\mcitedefaultseppunct}\relax
\EndOfBibitem
\bibitem[Milotti \latin{et~al.}(2022)Milotti, Berkmann, Laranjeira, Cui, Cao,
  Zhang, Kaiser, Yanagi, Melle-Franco, Shi, and Pichler]{Milotti22}
Milotti,~V.; Berkmann,~C.; Laranjeira,~J.; Cui,~W.; Cao,~K.; Zhang,~Y.;
  Kaiser,~U.; Yanagi,~K.; Melle-Franco,~M.; Shi,~L.; Pichler,~T. Unravelling
  the complete {Raman} response of graphene nanoribbons discerning the
  signature of edge passivation. \emph{Small Methods} \textbf{2022}, \emph{6},
  2200110--1--8\relax
\mciteBstWouldAddEndPuncttrue
\mciteSetBstMidEndSepPunct{\mcitedefaultmidpunct}
{\mcitedefaultendpunct}{\mcitedefaultseppunct}\relax
\EndOfBibitem
\bibitem[Talirz \latin{et~al.}(2019)Talirz, S{\"o}de, Kawai, Ruffieux, Meyer,
  Feng, M{\"u}llen, Fasel, Pignedoli, and Passerone]{Talirz19}
Talirz,~L.; S{\"o}de,~H.; Kawai,~S.; Ruffieux,~P.; Meyer,~E.; Feng,~X.;
  M{\"u}llen,~K.; Fasel,~R.; Pignedoli,~C.~A.; Passerone,~D. Band gap of
  atomically precise graphene nanoribbons as a function of ribbon length and
  termination. \emph{ChemPhysChem} \textbf{2019}, \emph{20}, 2348--2353\relax
\mciteBstWouldAddEndPuncttrue
\mciteSetBstMidEndSepPunct{\mcitedefaultmidpunct}
{\mcitedefaultendpunct}{\mcitedefaultseppunct}\relax
\EndOfBibitem
\bibitem[Sa{\"e}ki(1961)]{Saeki61}
Sa{\"e}ki,~S. The {Raman} spectra of 1,3,5-trichlorobenzene and
  1,3,5-trichlorobenzene-d$_3$. \emph{Bull. Chem. Soc. Jpn.} \textbf{1961},
  \emph{34}, 1851--1858\relax
\mciteBstWouldAddEndPuncttrue
\mciteSetBstMidEndSepPunct{\mcitedefaultmidpunct}
{\mcitedefaultendpunct}{\mcitedefaultseppunct}\relax
\EndOfBibitem
\bibitem[Tschannen \latin{et~al.}(2022)Tschannen, Vasconcelos, and
  Novotny]{Tschannen22}
Tschannen,~C.~D.; Vasconcelos,~T.~L.; Novotny,~L. Tip-enhanced Raman
  spectroscopy of confined carbon chains. \emph{J. Chem. Phys.} \textbf{2022},
  \emph{156}, 044203\relax
\mciteBstWouldAddEndPuncttrue
\mciteSetBstMidEndSepPunct{\mcitedefaultmidpunct}
{\mcitedefaultendpunct}{\mcitedefaultseppunct}\relax
\EndOfBibitem
\end{mcitethebibliography}

\providecommand{\latin}[1]{#1}
\makeatletter
\providecommand{\doi}
  {\begingroup\let\do\@makeother\dospecials
  \catcode`\{=1 \catcode`\}=2 \doi@aux}
\providecommand{\doi@aux}[1]{\endgroup\texttt{#1}}
\makeatother
\providecommand*\mcitethebibliography{\thebibliography}
\csname @ifundefined\endcsname{endmcitethebibliography}
  {\let\endmcitethebibliography\endthebibliography}{}

\end{document}